# Smart Foams: Switching Reversibly between Ultrastable and Unstable Foams

*Anne-Laure Fameau, Arnaud Saint-Jalmes, Fabrice Cousin, Bérénice Houinsou Houssou, François Boué, Bruno Novales, Laurence Navailles, Frédéric Nallet, Cédric Gaillard, Jean-Paul Douliez* \*

*((Dedication----optional))*

From materials chemistry to soft matter and daily life, foams are edifices of outstanding importance. They can be used as sacrificial templates for the production of macroporous materials [1, 2], in food and cosmetics or for a large variety of other applications [3-6]. Although a strong effort has been made for understanding the aging of foams[7] and consequently for increasing their stability and properties via their chemical formulation [8-10], there still does not exist a system which would exhibit high foamability in parallel with drastic reductions of the aging processes, leading to huge lifetimes. It must be mentioned however that the stability can be increased alternatively but by using gelator within the foam stock solution [11]. An o*ptimal* foam would be made from solutions of foam stabilizers which could move rapidly and readily at the interface (foamability) and subsequently produce an irreversibly adsorbed elastic layer at that interface, resisting to compression and thus limiting film breaking and gas diffusion (coalescence and coarsening). In addition, those ideal foam stabilizers should also contribute to limit the drainage in the liquid channels which form a continuous network between the bubbles, the so called Plateau borders, kinetically increasing the foam stability. Up to now, it remains difficult to conciliate both the foamability and the resultant foam stability, as optimizing one often reduces the other. For instance, partially hydrophobic solid particles allow the formation of very long-living foams [12-15] (over weeks) by forming a very solid layer at the interface, which annihilate the foam coarsening. The liquid drainage is usually not stopped in these solid-stabilized foams, though adding more hydrophilic particles could limit the liquid drainage via their accumulation in the Plateau borders [16]. However, these solutions of solid particles yield the production of a limited amount of foam, i.e., they exhibit poor foamability. This is mainly because the adsorption dynamic of those solid particles is slow and that the adsorption barriers are high. Solid particles also present the strong inconvenient that they aggregate in water and the foaming stock solution is not stable in time. By contrast, low molecular weight surfactants can more readily adsorb at the interface. However, they do not yield sufficiently solid film and film breaking or coarsening may occur more easily. For instance, SDS, CTAB or lauryl ether sulphate, which are the most common used surfactants have a good foamability, but produce foams with a stability that does not exceed a few tens of minutes. Protein solutions provide foams in between these two extremes, but often with still low foamability and aging [17-20]. Thus, it remains to find a simple system providing both high foamability and arrest of aging. Moreover, complementary very attractive feature of foams would be the possibility of tuning the foam stabilisation/destabilisation by a simple external trigger. Indeed, in some cases, good foam stability is initially required but the foam should be finally voluntarily destructed. This is the case for instance for recovering radioactive materials or in various washing processes. The first attempt to produce responsive foams have been realized using latex particles [21] increasing the pH by dropping diluted NaOH on the foam. However, such a drastic destabilizing procedure has the strong disadvantage that foams cannot be longer re-formed, except if one further re-decreases the pH of the solution what may not be satisfactory in many applications. Tuning the foamability (but not the foam stability) has been achieved by the temperature or UV irradiation [22]. If some results start to be collected on responsive self assemblies made of surfactants or gels (with reversible effects) [23], almost nothing has been found yet with foams.

Here, we report results on a simple and sustainable system which gathers a good foamability and an outstanding foam stability which can then be readily tuned to weak foam stability by changing the polymorphism of the system upon heating. To achieve this goal, we use as foam stabilizer a fatty acid system made of the 12 hydroxy stearic acid (12-HSA). Fatty acids exhibit low surface tension [24, 25] and have promising surface properties [26]. Moreover, the 12-HSA is known to form highly elastic layers at an air/water interface [25] (and even crystalline structures at high surface concentrations) because of strong intermolecular interactions induced by hydrogen bonding between hydroxyl groups. Besides, fatty acids are the model systems for investigating the properties of insoluble Langmuir interfacial

[*] ((A.L Fameau, B. Houinsou Houssou, Dr. B. Novales, Dr. C. Gaillard, Dr. J.P. Douliez))
((Biopolymères Interactions Assemblages))
((INRA))
((Rue de la Géraudière, 44316 NANTES (France)))
Fax: (+33)240-675-084
E-mail: jean-paul.douliez@nantes.inra.fr

((Dr. A. Saint-Jalmes))
((Institut Physique de Rennes))
((Université Rennes 1))
((263 avenue du Général Leclerc, 35042 RENNES (France)))

((Dr. F. Cousin, Dr. F. Boué))
((Laboratoire Léon Brillouin))
((CEA Saclay))
((Bat.563, 91191 GIF SUR YVETTE (France)))

((Dr. L. Navailles, Pr. F. Nallet))
((Centre de Recherche Paul Pascal))
((CNRS))
((115 Avenue Paul Schweitzer, 33600 PESSAC (France)))

[**] ((A.L Fameau would like to thank l'INRA and le CEA for the allocation of her Ph. D. grant. We thank the LLB for the SANS experiments. The access to the NMR facilities of the BIBS platform (Biopolymères Interactions Biologie Structurale) of INRA Angers -Nantes was greatly appreciated by the authors. We would like to thank Janine Emile (IPR, Rennes, France) for her precious help and powerful discussions during the thin film balance experiments))

Supporting information for this article is available on the WWW under http://www.angewandte.org or from the author.

monolayers [27, 28]. The use of fatty acids would allow us to significantly increase the foam stability by limiting film breaking and gas diffusion. However, the main problem is that fatty acids are insoluble or crystallise in water hampering their use as surface active agents [26]. To tackle this problem, one of our major achievement was to successfully disperse 12-HSA in water using an organic counter-ion (ethanolamine or hexanolamine): under such conditions, one obtains self assembled multilayer tubes of 10 μm length and 600 nm diameter [29, 30]. In solution, those tubes melt into micelles at a temperature which depends on the nature of the counter-ion [31]. Hence, we show here that the melting of tubes under micelles upon heating is accompanied by a complete destruction of the foam. This allows us tuning the foam stability by modifying the temperature. We first describe the foamability and the outstanding foam stability of that system at room temperature, establishing the link with the supramolecular assembly (tubes) in water. We further show how to destabilize that foam by simply heating and how it is related at local scale to the transition from tubes to micelles.

We produced foams using the multilamellar tubular system (figure 1) either by hand-shaking (HS-foams) or by using a foamscan ® apparatus (F-foams). The stability of HS-foams was evaluated over 6 months by comparing pictures taken at different times (Figure 2a). Foams were shown to be outstandingly stable (over weeks) suggesting us to describe them as 'ultrastable' foams. To our knowledge, such foam lifetimes obtained using low molecular weight amphiphiles are unprecedented and have only been observed up to now, when using solid particles [12, 13] or xanthane gelified foams stabilised by proteins [11].

The foaming properties were further studied on a shorter period of time, using F-foams (figure 2b). Producing foams by this procedure provides a *quantitative* description of that edifice in terms of how much foam volume is produced (foamability), how wet it is initially and how this foam ages. The foamability was found to be optimal since all the gas was encapsulated in the foam at the end of bubbling (see details in supporting information). The foam also entrapped large amount of liquid: at the end of bubbling, the average liquid fraction was around 20%, which corresponds to relatively wet foam, meaning that the bubble size has remained small and constant during the production (figure 2c). This shows that no bubble coalescence occurs, even at short times during bubbling. Interfaces are thus optimally and extremely rapidly covered as in the case of other low molecular weight surfactants. Once the foam is produced, its high stability was confirmed because the foam volume remained constant over the period studied and the bubble size did not change. However, the foam drained within 30 minutes (figure 2c). It is worth noting that it remains some liquid within the structure: about 5% in liquid fraction in average over the foam column. Then, these foams are mainly optimized in terms of no film breaking nor coarsening, but less efficiently from drainage reduction.

At this stage, one needs to determine the origins of this outstanding stability, and how this is connected to the supramolecular assemblies. The remaining presence of tubes in the foam was first confirmed by confocal microscopy using a hydrophobic fluorescent probe inserted into the multilamellar tubes. One clearly observed the rod like structures between the air bubbles (figure 3a and b). This shows that although some tubes unfolded upon foam formation (shearing) yielding free monomers which moved at the interface, they are still present in the foam liquid channels. Here, it must be noted that the length of the tubes is reduced by a factor of around two in the foam compared to the bulk solution (see figure 3, a and b). This reduction probably occurred during foam formation.

The structure of the supramolecular assembly of fatty acids at the local scale within the foam was probed by Small Angle Neutron Scattering (SANS). The experiments were carried on the foam and the results were compared with those obtained for the initial tube solution (stock solution) and for the liquid solution recovered at the bottom on the foam after drainage (drained solution), figure 3c). The three azimutally-averaged spectra were clearly similar. They show Bragg peaks at the same Q-position. Those peaks stand for the multilamellar arrangement of the fatty acid bilayers within the tubes [29] and confirm the presence of tubes in the foams. The only difference in the case of the foam is the $Q^{-4}$ scattering decay at low Q which accounts for the Porod surface scattering of the large bubbles of the foam. Although their length may have changed in the foam, the tubes are still composed of multilamellar fatty acid bilayers exhibiting similar properties than in the stock solution.

An other important information on the foam structure is at local scale in the thin lamella separating two bubbles which are in contact. Figuring out the mechanisms governing the stability of this thin lamella is crucial for explaining why bubbles coalesce or not and how gas diffuses from small to large bubbles (coarsening). For this, we have performed experiments on a single film with a "thin film balance" apparatus [32](see supporting information, figure S2). One observed in the single film a thin central part, with thickness of a few tens of nm, what is then called a "black film" [32] and implies that tubes are expelled from this film. That thin central part also exhibits a small area and is surrounded by a large meniscus much extended than what is usually found for other surfactant foams (see the details in the S2 picture). This shows that tubes are collected and jammed within that surrounding meniscus. By analogy in the foam, we expect the tubes to be expelled in the Plateau borders and absent from the lamella separating the bubbles. This specific structure of the lamella separating the bubble turns out to be quite efficient for the huge reduction of the foam aging. First, the thin central part of the lamella is made of two repulsing interfacial layers on which only free monomers of 12-HSA are adsorbed. These layers are most likely within a condensed-like state with the alkyl chains in close packing. It is well known that the presence of fatty acids adsorbed at an interface provide interfacial layers of high dilational moduli E reaching easily hundreds of mN/m [25, 27, 28] as the molecules are insoluble. Moreover, it is also known that hydroxylated fatty acids can pack at interface even more efficiently than usual fatty acids [25]. Such high dilational moduli (or low compressibility) is indeed the main reason why coarsening is almost stopped: the Gibbs criterion states that coarsening stops if $E > \gamma$, where $\gamma$ is the surface tension. In fact, if the bubble interfaces cannot be continuously compressed, the smallest bubble cannot vanish towards the largest ones, and coarsening is blocked [4, 6, 7].

Secondly, coarsening and lamella ruptures are also limited because the area of the thin contact part between bubbles (through which most of the gas diffusion occurs) is small, due to the presence of the tubes in the surrounding menisci which make them thicker and wider than usual. Adding the fact that the tubes can be jammed within the Plateau borders (at least, at the latest stages of drainage when the borders are already well shrunk) completes the picture which explains rather well the surprising long lifetime of these foams. We have thus here the first system which foams easily and which does not coarsen nor collapse because of an optimal arrangement of monomers and tubes within the foam structure.

Foams were further heated and their stability was monitored. It must be recalled that the multilamellar tubes melt into micelles at a temperature denoted Tm which depends on the nature of the counter-ion [31]. HS-foams made with the ethanolamine salt (Tm = 70°C) were still remarkably stable up to two weeks at 60°C. A similar high stability was observed for the hexanolamine salt (Tm = 60°C) at 50°C (see supporting information, S3). That high stability was quantitatively evidenced below Tm on the F-Foams (figure 4a). SANS data at those different temperatures were also similar in the foam, the stock solution and the drained solution (see supporting information S4). This shows that at all temperatures below Tm, tubes are always present in foams and exhibit the same behaviour as in bulk solution. However, at a temperature above that of the melting of the tubes, a foam could still be produced but was no longer stable as the foam volume quickly decreased with time (see figure 4a). Destabilisation occurred in only 15 minutes as observed on the HS-foams (using the hexanolamine salt, Tm = 60°C) placed in an oven at 60°C (see pictures, supporting information S5). This effect is dramatic since the destabilisation passes from weeks to few minutes in only few degrees. This event occurs exactly at the tube/micelle transition in both foams made of the ethanolamine and hexanolamine salts. The SANS data above that transition show that micelles (which have replaced tubes) are indeed present in solution but also in the foam (see figure 4b).

This offers us a versatile and simple way to produce temperature tuneable foams. The triggering of the foam stability by the temperature is illustrated in figure 5. An initial stable foam made with the hexanolamine salt (Tm = 60°C) at 20°C has a given constant foam volume which markedly decreases when the temperature is raised up to 60°C (figure 5a). Obviously, the foam is dramatically destabilized at that temperature in agreement with our previous experiments (figure 4a). Remarkably, decreasing back the temperature to 20°C is accompanied by a sudden stop of the decrease of the foam volume which becomes constant again as a function of time. If the foam is initially produced at T>Tm, it is not stable as evidenced by the quick decrease of the foam volume (figure 5b). However, here again lowering the temperature below Tm, i.e. 55°C, a temperature at which the tubes are reformed, enables to recover a very stable foam. Additional curves are illustrated in the supporting information S6. Altogether, this suggests that tubes reformed in the foam upon cooling, as observed in the bulk when crossing the tube-micelle transition, allowing recovering the stabilisation of the foam. This almost instantaneous phenomenon yields a complete reversibility to the system. Moreover, the foam can also be regenerated (when it has been destructed at high temperature) by a simple re-injection of gas in the column below the temperature at which tubes melt into micelles. Here, we can only speculate on the origins of foam destabilization. Many effects which can lead to lamella rupture and foam collapse can occur, as listed below. Lamella and foams are fragile and brittle edifices especially if the interfacial layers are solid-like as expected here. So when the temperature is raised, interfacial flows, changes in the shape of the lamellas and menisci around them must be triggered easily leading to the lamella rupture. We also believe that above the tube/micelle transition, the interfacial layers become softer and less condensed also because fatty acids become more soluble at high temperature. Moreover, the exchange of monomers between the interface and the micelles is expected to be faster than between the interface and the tubes.

Then, we have shown that in terms of foam stability, this system made of multilamellar tubes of 12-hydroxystearic acid combines the advantages of both the solid particles and the amphiphiles, since those fatty acids readily and quickly go at the interface but also produce solid layers which cannot be indefinitely compressed. In addition, the tubes lead to large menisci surrounding the lamella, and reduce the drainage flows. Of particular interest is that the reversible phase transition of tubes into micelles upon heating yields reversible fast destabilisation. This system is unique compared to other amphiphile-forming tubes which generally melt into vesicles [33]. Moreover, the multilamellar tubes can survive to various changes of physico-chemical parameters [34]. 12-hydroxy stearic acid is a generic molecule that results from the hydrogenation of a sustainable material, i.e., ricinoleic acid. It is available in large amount at low cost. Because the tube/micelle transition can be tuned by the nature of the counter-ion [31], one could design similar systems, leading to foams being destabilized at any desired temperature.

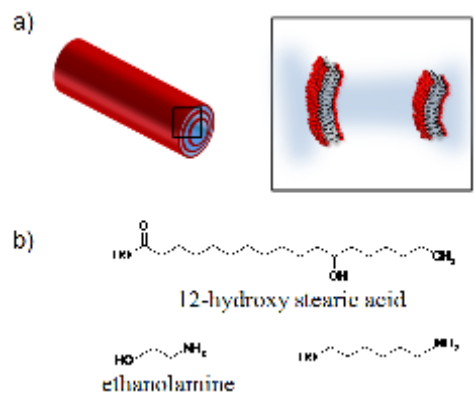

*Figure 1.* a) Schematic representation of the multilayer tubes and b) chemical structure of the 12-hydroxy stearic acid and the counter-ions used.

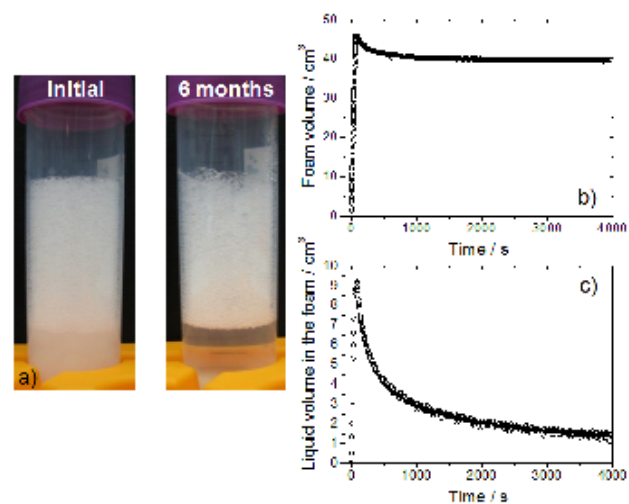

*Figure 2.* (a) Photos of HS-foams taken at different times in the system 12-HAS/ethanolamine at 40 mg/mL fatty acids. The turbid tubular solution which has drained is visible in the lower part of the sample tube. Clearly, images at 6 months interval are rather similar without noticeable change of the volume showing that our foams are 'ultrastable'. The sole difference being that tubes sediment in the bottom of the tube after 6 months. (b) Foaming properties as recorded by the foamscan ® apparatus in the system 12-HSA/ethanolamine at 10 mg/mL fatty acids. Using this instrument, F-foams are produced by bubbling gas (flow rate= 35 mL/min) in the solution containing tubes via a porous glass filter (pore size 10-14 μm) creating bubbles of diameter ranging between 50 to 100 microns. A maximal foam volume is fixed and reached at 78 s. Then, the gas flow is stopped and the foam volume is determined by a CCD camera. (c) The evolution of the liquid volume in the foam determined by conductivity measurement as a function of time is also shown. That graph illustrates the drainage occurring in the foam upon aging.

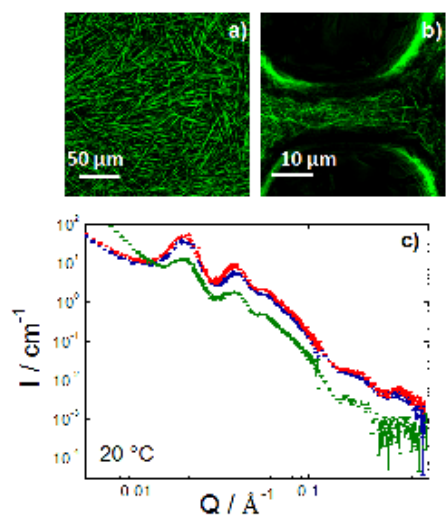

*Figure 3.* Confocal microscopy images of a bulk solution of tubes (a) and of a foam (b) which was deposited on glass lamellae and further covered with a glass plate using a spacer ensuring a fixed volume of 25 µL. (c) SANS data recorded for the foam (●), the stock solution (●) and the drained solution (●). In the case of foams, it was first checked by contrast variation experiments that the recorded signal corresponded to SANS scattering and not to neutron reflectivity [35] (see Supporting Information, figure S1).

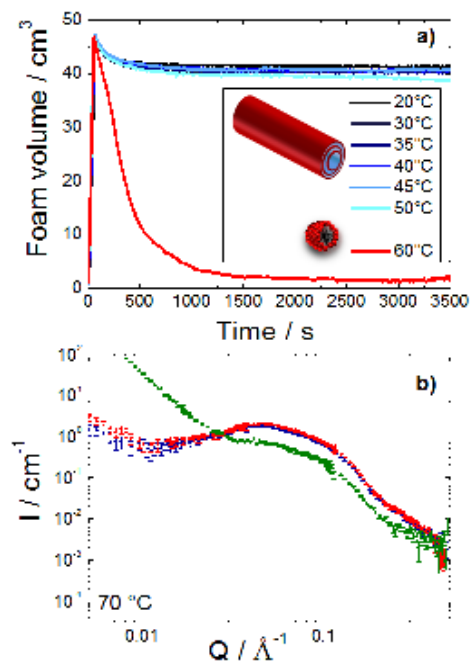

*Figure 4.* Evolution of the foam volume as a function of time in F-foams at different temperatures for the 12-HSA/hexanolamine system. SANS data at 70°C (in the 12-HSA/ethanolamine system) showing that micelles are formed at that temperature in both the foam (●) and the aqueous solutions (drained (●) and stock (●)).

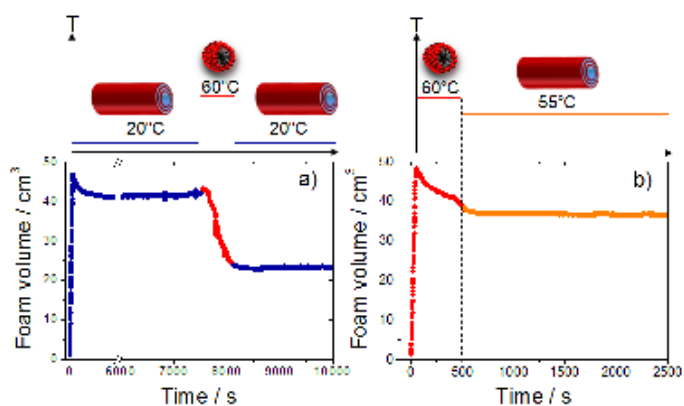

*Figure 5.* Two examples showing the evolution of the foam volume as a function of time and temperature (in the 12-HSA/hexanolamine system). On the top of each graph is shown the schematic representation of the supramolecular assemblies (not at scale) present in solution and in the foam as a function of the temperature.

**Entry for the Table of Contents**

Anne-Laure Fameau, Arnaud Saint-Jalmes, Fabrice Cousin, Bérénice Houinsou Houssou, François Boué, Bruno Novales, Laurence Navailles, Frédéric Nallet, Cédric Gaillard, Jean-Paul Douliez.

**Smart Foams: Switching Reversibly between Ultrastable and Unstable Foams**

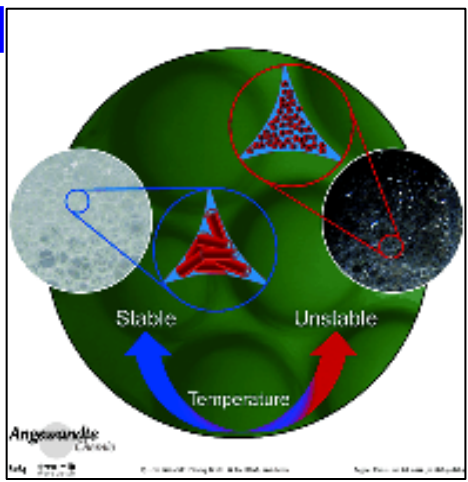

Smart foams: Ultrastable foams with an optimal foamability have been obtained using hydroxyl fatty acids tubes. The stabilization results from the adsorption of monomers at the air-water interface preventing coalescence and coarsening and from the presence of tubes in the Plateau borders limiting the drainage. Upon heating, tubes transit to micelles, which induces foam destabilization. Such foams are thus the first to have a temperature tunable stability.